\documentstyle[12pt]{article}

\begin{document}

\pagestyle{empty}

\noindent {\small USC-97/HEP-B1\hfill \hfill hep-th/9703060}\newline
{\small LPTENS-97/09\hfill CERN-TH/97-36}

{\vskip 0.7cm}

\begin{center}
{\LARGE Theories with Two Times } \\[0pt]

{\vskip 0.5cm}

{\bf Itzhak Bars}$^a${\bf \ and Costas Kounnas}{$^b$} {\Large \ \\[0pt]
}

{\vskip 0.4cm}

{\bf TH Division, CERN, CH-1211 Geneva 23, Switzerland}

{\vskip 0.5cm}

{\bf ABSTRACT}
\end{center}

\noindent
{\small General considerations on the unification of A-type and B-type
supersymmetries in the context of interacting p-branes strongly suggest that
the signature of spacetime includes two timelike dimensions. This leads to
the puzzle of how ordinary physics with a single timelike dimension emerges.
In this letter we suggest that the two timelike dimensions could be real,
and belong to two physical sectors of a single theory each containing its
own timelike dimension. Effectively there is a single time evolution
parameter. We substantiate this idea by constructing certain actions for
interacting p-branes with signature }$\left( n,2\right) ${\small \ that have
gauge symmetries and constraints appropriate for a physical interpretation
with no ghosts. In combination with related ideas and general constraints in
S-theory, we are led to a cosmological scenario in which, after a phase
transition, the extra timelike dimension becomes part of the compactified
universe residing inside microscopic matter. The internal space, whose
geometry is expected to determine the flavor quantum numbers of low energy
matter, thus acquires a Minkowski signature. The formalism meshes naturally
with a new supersymmetry in the context of field theory that we suggested in
an earlier paper. The structure of this supersymmetry gives rise to a new
Kaluza-Klein type mechanism for determining the quantum numbers of low
energy families, thus suggesting that the extra timelike dimension would be
taken into account in understanding the Standard Model of particle physics.}
\bigskip \bigskip

\vfill
\hrule width 6.7cm \vskip 2mm

{$^a$ {\small On sabbatical leave from the Department of Physics and
Astronomy, University of Southern California, Los Angeles, CA 90089-0484,
USA.}}

{$^b$ {\small On leave from Ecole Normale Sup\'{e}rieure, 24 rue Lhomond,
F-75231, Paris, Cedex 05, France. }}

\vfill\eject

\setcounter{page}1\pagestyle{plain}

\medskip

\section{Hidden dimensions from generalized SUSY}

Duality properties of string theory have shown that the basic theory behind
string theory contains open and closed super p-branes interacting with each
other. The fundamental form of the theory is unknown, however an overall
property of the theory is the presence of A and B sectors related to each
other by duality, such that in each sector a superalgebra with as many as 32
real supercharges and as many as 528 bosonic charges govern general
properties of the interactions (this includes the heterotic sectors with 16
supercharges). The 32 supercharges correspond to the maximum number N=8
conserved Majorana supercharges in four flat dimensions. The 528 bosons
correspond to all possible open and closed p-brane sources that can couple
in all closed/open sectors. Some or all of these fermionic/bosonic charges
may vanish in different sectors, including heterotic sectors. Thus, whatever
the form of the theory, in its flat limit it must obey a generalized form of
the superalgebra. The maximum amount of information is obtained when all 32
supercharges are active.

By examining various reclassification schemes of the A,B superalgebras it
was recognized that quantum numbers associated with hidden dimensions may be
attached to the N=8 labels in several ways \cite{ibtokyo}-\cite{ibstheory}.
In four dimensions one may distinguish 3 classification schemes of the
internal N=8 labels that {\it are not contained in each other}. Namely N=8
corresponds to 8$\oplus $8$^{*}$ of SU(8), or to the spinors 8$_{+}\oplus
8_{-}$ of SO(7,1), or to the spinors (4,2)$\oplus $(4$^{*},2)$ of SO(6)$
\times $SO(2,1). Obviously these groups are not contained in each other. We
refer to these as the duality basis, the A basis and the B basis
respectively. Once the 32 spinors are classified, the 528 bosons obtain a
unique classification since they appear in the products of the spinors. The
transformations from one basis to another are related to duality
transformations. Each classification includes an SO(6) corresponding to 6
compactified dimensions already familiar from string theory. In the
references above it was argued that the meaning of SO(7,1) is 7 spacelike
and 1 timelike internal dimensions, and that the meaning of SO(6)$\times $
SO(2,1) is 8 spacelike and 1 timelike internal dimensions. The SU(8)
classification is useful for studying certain duality properties of the
theory but this basis obscures the other two classifications that give
information about hidden dimensions in the theory (recall examples where
electric-magnetic duality is obscured by Lorentz covariance and vice versa).

The three classifications are present in various dimensions. The content of
the highest hidden dimensions is clearly displayed by rewriting the A,B type
superalgebras directly in SO(10,2) and SO(9,1)$\times $SO(2,1) covariant
forms respectively, suggesting that altogether there may be as many as 13
dimensions \cite{ibstheory}. It was then noted that the A,B superalgebras
correspond to two distinct projections from a bigger superalgebra that has
64 fermions and 2080 bosons (=78+286+1716) which are covariant under
SO(11,2). This suggested that the underlying unified theory of p-branes may
be formulated in (11,2) dimensions. In recent work \cite{ibso22} it has been
argued that the projection down to the A,B sectors that contain the two
types 32$_{A,B}$ supercharges may be implemented in a SO(11,2) covariant
form as a BPS\ condition. Furthermore, it was pointed out that the same 64
fermions and 2080 bosons (=364+1716) are actually covariant under SO(11,3)
and the A,B projections may be implemented with an SO(11,3) covariant BPS
condition. Therefore the fundamental supersymmetric theory may admit as many
as 11 spacelike and 3 timelike dimensions, and this number of dimensions may
be a limit due to supersymmetry.

If taken seriously this would suggest that at the fundamental level there
are p-branes $X^M(\tau ,\sigma _1,\cdots ,\sigma _p)$ where the target
spacetime index $M$ describes signature $\left( 11,2\right) $ or $\left(
11,3\right) $ in order to unify all sectors of the theory. For supersymmetry
one should also take the 64 spinorial fermions $\theta _\alpha (\tau ,\sigma
_1,\cdots ,\sigma _p),$ with appropriate gauged kappa supersymmetries. The
immediate issue that arises is how to construct a physical theory that
contains such extra timelike dimensions? In particular how would one avoid
the problems of two or more timelike dimensions, such as ghosts, causality,
etc., that have been noted over many past decades?

Evidently there should be enough gauge symmetry in the construction of an
action while preserving covariance. The gauge symmetry should lead to
appropriate covariant constraints that permit a physical interpretation
equivalent to a single time evolution parameter. This is the problem for
which we suggest a solution. In the next section a covariant gauge mechanism
is illustrated in an instructive toy model of interacting 0-branes, i.e.
particles. The simple mechanism that solves the problem is generalized to
strings and other p-branes in additional work elsewhere \cite{ibkounnas2}.
Our approach makes connections to other arguments and models about the
fundamental theory that also involve (10,2) signature \cite{ademo}-\cite
{tseytlin}, and suggests the natural setting that leads to such models as part
of a more complete system. The solution provides the proper interpretation
of our recent construction of field theoretic models with a new
supersymmetry. Combining these observations with general properties of the
superalgebra pointed out in the context of S-theory \cite{ibstheory}\cite
{sentropy} (in particular as related to black holes) lead us to a
cosmological scenario for the emergence of the standard universe with a
single time coordinate from a theory of p-branes with more than one timelike
dimensions. The extra timelike dimension(s) have consequences for the flavor
quantum numbers in low energy physics.

\section{Two particles}

Consider the world lines of two particles (zero branes) $x_1^\mu \left( \tau
\right)$, $x_2^\mu \left( \tau \right)$ with the following action

\begin{eqnarray}
S &=&S_1+S_2+S_{12}  \nonumber \\
S_n &=&\frac 12\int_0^Td\tau \,\left( e_n\dot{x}_n^2-\frac{m_n^2}{e_n}\right)
\\
S_{12} &=&\eta _{\mu \nu }\int_0^Td\tau \,\dot{x}_1^\mu A_1\int_0^Td\tau \, 
\dot{x}_2^\nu A_2  \nonumber
\end{eqnarray}
where the familiar $S_1(x_1^\mu ,e_1),\ S_2(x_2^\mu ,e_2)$ describe the free
motion of each particle
\footnote{
If one substitutes the solutions for $e_{1,2}$ in the actions, then $S_{1,2}$
take the more familiar form $m_{1,2}\int d\tau \sqrt{-\dot{x}_{1,2}^2}$.}. 
$S_{12}$ is an interaction involving the additional worldline fields $
A_1\left( \tau \right)$, $A_2\left( \tau \right) $. This action is
invariant under separate global spacetime translations of both $x_1^\mu $
and $x_2^\mu $ and {\it common} global rotations of all spacetime
directions. The signature of the flat spacetime metric $\eta_{\mu \nu}$ {\it 
will not be a priori specified}, and it will be shown that the solution
space {\it requires} a signature with at least two timelike dimensions. It
will also be shown that the propagation of each particle can depend on only
one linear combination of these timelike dimensions, while the other
particle propagates in the orthogonal timelike dimension, thus each particle
being consistent with propagation with a single time coordinate. This will
be interpreted physically in a cosmological scenario.

The action is invariant under two {\it independent} $\tau $
reparametrizations of the fields [$x_n^\mu \left( \tau \right),e_n\left(
\tau \right),A_n\left( \tau \right) $] for each $n=1,2$ which allow the two
gauge choices $e_n=1$ eventually. There are also two independent gauge
invariances associated with the fields $A_n$ given by 
\begin{eqnarray}
\delta x_1^\mu  &=&\lambda _2^\mu \Lambda _1\left( \tau \right) ,\quad
\delta A_1=-\left( e_1+\frac{\lambda _2^2A_1}{\lambda _2\cdot \dot{x}_1}
\right) \partial _\tau \Lambda _1,  \label{gauge} \\
\delta x_2^\mu  &=&\lambda _1^\mu \Lambda _2\left( \tau \right) ,\quad
\delta A_2=-\left( e_2+\frac{\lambda _1^2A_2}{\lambda _1\cdot \dot{x}_2}
\right) \partial _\tau \Lambda _2.  \nonumber
\end{eqnarray}
where we have defined the dynamically determined constant vectors 
\begin{equation}
\lambda _1^\mu =\int_0^Td\tau \,\dot{x}_1^\mu A_1,\quad \lambda _2^\mu
=\int_0^Td\tau \,\dot{x}_2^\mu A_2.
\end{equation}
Note that the fields labeled by $n=1,2$ are mixed by this gauge
transformation since $\lambda _2^\mu $ appears in the transformations of
particle \#1, and vice-versa. It is due to these gauge invariances, which
can remove one component from each $x_n^\mu \left( \tau \right) ,$ that
effectively there is a single time coordinate. However, as we will see
shortly, the dynamics will determine the two $\lambda _n^\mu $ to be
orthogonal and timelike (or lightlike in the case of vanishing masses) thus
requiring a signature $\left( n,2\right) $ with two or more timelike
dimensions.

Each particle moves in a dynamically defined background $\lambda _n^\mu $
provided by the other particle. For particle \#1 moving in the background
of particle \#2 the effective action is 
\begin{equation}
S_{1(2)}=\frac 12\int_0^Td\tau \,\left( e_1\dot{x}_1^2-\frac{m_1^2}{e_1}
\right) +\lambda _{2\mu }\int_0^Td\tau \,\dot{x}_1^\mu A_1
\end{equation}
The last term may be interpreted as an interaction of the particle with an
``electromagnetic'' potential of a specific form 
\begin{equation}
A_\mu \left( \tau \right) =\lambda _{2\mu }A_1\left( \tau \right) ,
\end{equation}
where $\lambda _{2\mu }$ plays the role of a polarization vector. The
canonical momentum is conserved due to the equation of motion 
\begin{equation}
p_1^\mu =e_1\left( \tau \right) \,\dot{x}_1^\mu \left( \tau \right) +\lambda
_2^\mu A_1\left( \tau \right) ,\quad \partial _\tau p_1^\mu =0.
\end{equation}
The vector $\lambda _1^\mu $ is rewritten in terms of canonical variables 
\begin{equation}
\lambda _1^\mu =\int_0^Td\tau \,\dot{x}_1^\mu A_1=\int_0^Td\tau \,\hat{p}
_1^\mu A_1/e_1
\end{equation}
where 
\begin{equation}
\hat{p}_1^\mu =p_1^\mu -\lambda _2^\mu A_1
\end{equation}
satisfies constraints and equations of motion that follow from the action 
\begin{equation}
\lambda _2\cdot \hat{p}_1=0,\quad \hat{p}_1^2+m_1^2=0,\quad \partial _\tau 
\hat{p}_1^\mu =0.
\end{equation}
The first constraint shows (without using the equation of motion or special
gauges) that $\lambda _1^\mu $ is orthogonal to $\lambda _2^\mu $ 
\begin{equation}
\lambda _1\cdot \lambda _2=0.
\end{equation}
The equation $\partial _\tau \hat{p}_1^\mu =0$ follows from $\partial _\tau 
{p}_1^\mu =0$ and the first constraint, because $A_1=p\cdot \lambda
_2/\lambda _2^2$ is a constant provided the denominator does not vanish. If $
\lambda _2^2=0$ then $A_1\left( \tau \right) $ is undetermined, but it can
be chosen to be a constant by using the gauge freedom  $\delta x_1^\mu
=\lambda _2^\mu \Lambda _1\left( \tau \right) ,\,\,\delta A_1=-e_1\partial
_\tau \Lambda _1$. Either way $\partial _\tau \hat{p}_1^\mu =0$ is
satisfied. Since $\hat{p}_1^\mu $ is conserved then $\lambda _1^\mu $ is
proportional to it 
\begin{equation}
\lambda _1^\mu =\hat{p}_1^\mu \ \ \int_0^Td\tau \,A_1/e_1
\end{equation}
The remaining constraint (mass shell condition) requires that $\lambda
_1^\mu $ is timelike for $m_1\neq 0$ or lightlike for $m_1=0.$ Similar
arguments hold for $\lambda _2^\mu .$ Choosing the gauges $e_n=1,$ the
overall physical solution is given by 
\begin{eqnarray}
x_1^\mu \left( \tau \right) =\hat{p}_1^\mu \tau +q_1^\mu ,\quad x_2^\mu
\left( \tau \right) =\hat{p}_2^\mu \tau +q_2^\mu  \\
\hat{p}_n^2 =-m_n^2,\quad \hat{p}_1\cdot \hat{p}_2=0,\quad \lambda _n^\mu =
\hat{p}_n^\mu A_nT,
\end{eqnarray}
where the $A_n$ are constants.

According to the solution space the particles move freely except for the
constraint that the momenta $\hat{p}_1^\mu ,\hat{p}_2^\mu $ must be
orthogonal. Furthermore, in the massive case, since both vectors are
timelike the constraints cannot be satisfied with only one timelike
dimension. One must have at least two timelike dimensions, therefore we take
the signature for the spacetime index $\mu $ as ($n,2).$ Yet, effectively
each particle moves in a subspace of signature $\left( n,1\right) .$ For
example in the rest frame of particle \#2 one can write the momenta in the
form 
\begin{equation}
\hat{p}_2^\mu =\left( m_2,0;{\bf 0}\right) ,\,\quad \,\hat{p}_1^\mu =\left(
0,\sqrt{m_1^2+{\bf p}_1^2};{\bf p}_1\right) 
\end{equation}
showing that the energy-momentum relations are as usual. In addition, note
that there is in fact only one propertime parameter $\tau $ to describe the
time evolution (because it is possible to choose both $e_n=1$).

If one of the particles is massless, say $m_2=0,$ then $\lambda _2^\mu \sim 
\hat{p}_2^\mu $ \thinspace is lightlike. As mentioned above, one may use the
gauge freedom to fix $A_1.$ Thus, one may choose $A_1$ to be just the
required constant so that $\hat{p}_1^\mu =p_1^\mu -\lambda _2^\mu A_1$ has
no components along $\hat{p}_2^\mu .$ Therefore, the solution space is again
as above, but now $\hat{p}_1^\mu $ has two less components, one timelike and
one spacelike, since it cannot point along the lightlike vector $\hat{p}
_2^\mu .$ So it propagates in a space of signature $(n-1,1)$. For example,
in a special frame one may write 
\begin{equation}
\hat{p}_2^\mu =\left( \left| p_2\right| ,0;p_2,\vec{0}\right) ,\quad \ \hat{p
}_1^\mu =\left( 0,\sqrt{m_1^2+\vec{p}_1^2};0,\vec{p}_1\right) .
\label{massless}
\end{equation}
We emphasize that one may change
\begin{equation}
\hat{p}_1^\mu \rightarrow \hat{p}_1^\mu +\alpha \hat{p}_2^\mu 
\label{freedom}
\end{equation}
for any constant $\alpha ,$ but this is a gauge freedom. By choosing $\alpha
=0$ we keep the interpretation that each particle uses only one timelike
coordinate.

If both particles are massless $m_1,m_2=0,$ the arguments above may be
repeated to find out that the $A_n$ may be chosen as constants just as to
insure that $\hat{p}_n^\mu $ are time independent, lightlike and not
parallel to each other (the last possibility is a gauge choice). In a
special frame they may be written as in eq.(\ref{massless})  in the
limit $m_1=0.$

The lightlike case for $\hat{p}_2^\mu $ is reminiscent of a lightlike vector
introduced in the context of several attempts for constructing theories with
signature $\left( n,2\right) .$ These include heterotic (p,q) strings
involving left/right movers with N=2 supersymmetry \cite{oogurivafa}\cite
{kutmartinec}, F-theory \cite{vafa}, $\left( 10,2\right) $ super Yang-Mills 
\cite{sezgin}, $\left( 10,2\right) $ version of a matrix model for M-theory 
\cite{periwal}. We suggest that a natural interpretation of the lightlike
vector in these theories would be obtained by considering the type of system
discussed in this paper. Namely consider the presence of a 0-brane or a
p-brane \#2 which provides a background to the strings considered in these
models. Then in the sector of lightlike $\hat{p}_2^\mu $ the models would be
recovered, but there would be now the additional system \#2 to be taken into
account including all the allowed values and directions of the vector $\hat{
p }_2^\mu .$ This will remove the rigidity of the lightlike vector and
establish full SO$\left( n,2\right) $ covariance for the full model. Such
generalized models will be discussed in another paper \cite{ibkounnas2}.

The quantum theory of the two particle system can be written covariantly
with SO($n,2$) symmetry. One may start with naive quantization rules 
\begin{equation}
\left[ x_m^\mu ,\hat{p}_n^\nu \right] =i\eta ^{\mu \nu }\delta _{mn},\quad
\left[ x_m^\mu ,x_n^\nu \right] =\left[ \hat{p}_m^\mu ,\hat{p}_n^\nu \right]
=0
\end{equation}
and impose the constraints on the states. Since there are two particles, the
wavefunction in position space depends on both coordinates with $\left(
n,2\right) $ signature, and satisfies a Klein-Gordon type equation
corresponding to the constraints 
\begin{equation}
\left( \partial _n^2-m_n^2\right) \Phi \left( x_1^\mu ,x_2^\mu \right)
=0,\quad \partial _1\cdot \partial _2\,\,\Phi \left( x_1^\mu ,x_2^\mu
\right) =0.
\end{equation}
The general solution is the general superposition of a product of plane
waves with momenta that are on shell and orthogonal to each other 
\begin{eqnarray}
\Phi \left( x_1^\mu ,x_2^\mu \right)  &=&\int
d^{n+2}k_1\,d^{n+2}k_2\,\,\delta \left( k_1^2+m_1^2\right) \delta \left(
k_2^2+m_2^2\right) \delta \left( k_1\cdot k_2\right)   \\
&&\left[ a_+\left( k_1,k_2\right) e^{i\left(
k_1\cdot x_1+k_2\cdot x_2\right) }+ a_-\left( k_1,k_2\right) e^{i\left(
k_1\cdot x_1-k_2\cdot x_2\right) }+ c.c.\right] .  \nonumber
\end{eqnarray}
The delta functions impose the constraints. The coefficient $a_+\left( k_1,k_2\right)$ includes a theta function $\theta (k_1^0{k_{2}^{0'}}-{k_{1}^{0'}}k_2^0)$ while $a_-$ contains a theta function with the opposite argument. The sign of the argument of the theta function cannot be changed with SO$(n,2)$ transformations, therefore including the theta functions is the analog of imposing the positive energy condition. Hence the coefficients $a_{\pm}$ have the interpretation of probability amplitude for (particle \#1, particle \#2) and (particle \#1, antiparticle \#2) respectively. For their complex conjugates particle is interchanged with antiparticle. This interpretation of the coefficients is consistent with SO(n,2) covariance. 

If one or both particles are massless additional
conditions are needed to implement the gauge freedom of eq.(\ref{freedom}) to
insure that the timelike coordinates of the particles do not overlap. The
gauge freedom (for the case $m_2=0$) corresponds to translating $k_1^\mu
\rightarrow k_1^\mu +\alpha k_2^\mu $ and integrating over $\alpha $. This produces an additional delta function $\delta \left(
k_2\cdot x_1\right) $ to be inserted in the integral above. Similarly, if $
m_1$ is also zero there would be another constraint implemented by $\delta
\left( k_1\cdot x_2\right).$ Of course the constrained momenta point in all
possible directions in the spacetime of $\left( n,2\right) $ signature to be
able to recover all possible quantum states and full covariance under SO$
\left( n,2\right) $.
 
There are no ghosts in the spectrum of the first quantized theory. A field theory
that gives precisely only such solutions can be written down as in our
previous paper on a ``new supersymmetry'' \cite{superpv}. The field theory
that we proposed has higher derivative terms, but they are arranged in such
a way that ghost problems are avoided as demonstrated there. Also, in \cite
{superpv} the field theory corresponds to a compactified version of the two
particle problem, but this has been generalized to the general case in \cite
{ibkounnas3} with similar results.

\section{New Supersymmetry}

The supersymmetric version of the two particle field theory described above
may be constructed along the lines suggested in our recent paper \cite
{superpv}. For simplicity we will consider $N=1$ in $\left( n,2\right) $
dimensions. Let the supercharge be denoted by $Q_\alpha $. We had shown that
for a superalgebra of the form 
\begin{equation}
\left\{ Q_\alpha ,Q_\beta \right\} =\gamma _{\alpha \beta }^{MN}Z_{MN},\quad
Z^{MN}=\hat{p}_1^M\hat{p}_2^N-\hat{p}_1^N\hat{p}_2^M,  \label{special}
\end{equation}
one can construct representations and Lagrangians in a field theory with
bilocal fields $\Phi (x_1^\mu ,x_2^\mu )$. Actually we had considered a
compactified version of this form (see also below eq.(\ref{compactified}))
but the approach is generalizable to the present case \cite{ibkounnas3}. In
our previous paper a two particle interpretation was not given but the
formalism that was used was equivalent to it. With our current understanding
the two particle interpretation of this superalgebra is quite natural.

If we use the constraints on the momenta derived in the previous section
this superalgebra reduces to the standard one in a special frame of each
particle. For example consider the case of massive particle \#2. In its rest
frame its momentum points along the $0^{\prime }$ direction and the
superalgebra reduces to 
\begin{equation}
\left\{ Q_\alpha ,Q_\beta \right\} =2m_2\gamma _{\alpha \beta }^{0^{\prime
}\mu }Z_{0^{\prime }\mu }=2m_2\left( \gamma ^{0^{\prime }}\gamma ^\mu
\right) _{\alpha \beta }\,\,p_{1\mu }
\end{equation}
This is recognized as the one particle superalgebra for a spacetime with
signature $\left( n,1\right) $. If $n$ is even, and if the spinor index $
\alpha $ corresponds to a Weyl spinor for SO$\left( n,2\right) ,$ then $
Q_\alpha $ is an irreducible spinor in $\left( n,1\right) $ dimensions. If $n
$ is odd then $Q_\alpha $ splits into two spinors of opposite chirality in $
\left( n,1\right) $ dimensions.

For a massless particle \#2 in its special frame the $\gamma ^{0^{\prime }}$
would be replaced by the lightcone combination $\gamma ^{0^{\prime
}+1^{\prime }}$ and the superalgebra would become the one particle
superalgebra in a spacetime with signature $\left( n-1,1\right) .$ Half of
the original supergenerators $Q_\alpha $ vanish because they satisfy a BPS
condition. The other half is the spinor for SO$\left( n-1,1\right) $ that
forms the $N=1$ superalgebra in the lower dimension that describes the
motion of particle \#1 (reduced by one spacelike and one timelike
dimensions).

For fields $\Phi (x_1^\mu ,x_2^\mu )$ that describe all possible states
which are not necessarily in special frames the superalgebra (\ref{special})
becomes 
\begin{equation}
\left\{ Q_\alpha ,Q_\beta \right\} =-2\,\gamma _{\alpha \beta
}^{MN}\,\partial _{1M}\partial _{2N}  \label{derivatives}
\end{equation}
This is the form for which representations can be found as discussed in a
previous paper \cite{superpv} and a future one \cite{ibkounnas3}. Such
representations include bilocal bosonic and fermionic fields and generalize
the representations of standard supersymmetry.

The special forms of the superalgebra (\ref{special},\ref{derivatives}) were
first considered in the context of S-theory \cite{ibstheory}. Mass shell
conditions corresponding to BPS conditions which restricted $\hat{p}_n^M$
were considered, including conditions such as $\hat{p}_1\cdot \hat{p}_2=0$
as in the present paper. The form of $Z^{MN}$ was an anzats motivated by
certain applications of S-theory, but at the time it was not known why such
an algebra would arise. It was simply noted that it would be the minimal
algebra required in a $\left( 10,2\right) $ version of supergravity. A
compactified version, including central extensions, was also applied to the
problem of labeling black hole states and computing the black hole entropy 
\cite{sentropy}. The two particle constrained problem introduced in the
current paper provides a natural setting for this form of new supersymmetry
and all of its previous applications. As insisted in S-theory and in its
applications, the two momenta must be allowed to take all possible values in
the spacetime of signature $\left( n,2\right) $ in order to recover the full
SO$\left( n,2\right) $ covariance of the underlying theory. The two particle
interpretation demands this naturally.

It is interesting to point out that it was later noted that the same form
for the superalgebra appears in $\left( 10,2\right) $ versions of super
Yang-Mills theory \cite{sezgin}, $\left( 2,1\right) $ superstrings \cite
{martinec} and matrix models for M-theory \cite{periwal}. However, since in
those applications $\hat{p}_2^M$ was taken as a rigid lightlike vector, the
SO$\left( 10,2\right) $ covariance was broken. The SO$\left( 10,2\right) $
covariance would be valid in a larger theory in which the lightlike vector
is allowed to point in all possible directions \cite{ibst96}\cite{sentropy}.
This generalization is now justified by suggesting that those models can be
reconsidered and interpreted naturally in a setting analogous to the two
particle problem discussed in this paper and in \cite{ibkounnas2}.

\section{Cosmological scenario and low energy physics}

If at some deep level there is an extended universe that functions with two timelike dimensions, how
does it evolve to our present universe in which there is only one time? It is
common to assume that there was a phase transition during the early universe
era, such as the Big Bang, which triggered the expansion of some of the
dimensions while the others remained compact and small. In conjunction with this phase transition, a plausible scenario is that one of the
timelike dimensions goes along with the expanding universe and the other
goes along with the compactified one as described below. For instance, in the two particle model (or its generalization to p-branes) $p_1^\mu$ would be in four dimensions while $p_2^m$ would be in the compactified dimensions including the extra timelike dimension. Amusingly, such a picture seems to be consistent with several other observations and therefore deserves some study, even though the physical mechanism that triggers the Big Bang remains obscure.

Let us first connect it to some observations about black holes, since sometimes it is useful to think of black hole singularities as the opposite
of Big Bang type white hole singularities. In that case one may expect that
the extra compactified timelike and spacelike dimensions reside also inside black holes.
In fact, there is already a clue in favor of this picture in agreement
with the scenario we propose, which indicates that black holes do have information
about the hidden time and space coordinates. During the past year new
advances were made in computing the entropy, for certain stringy charged black holes, in
terms of D-branes \cite{stromvafa}. The result shows that the black hole
entropy is written in terms of the central extensions of the
superalgebra \cite{kallosh} which describe the wrappings of p-branes in internal spacelike
dimensions. By using the reclassification of the supercharges mentioned in the first section, it was also shown that the expression for the black hole entropy corresponds to evaluating it in the rest frame of an internal Lorentz space SO$(c+1,1)$ where 
$c+1$ is the number of compactified string coordinates $c$ plus the 11th
dimension, while the timelike direction corresponds to the 12th dimension.
The same black hole entropy was then written in the general SO$(c+1,1)$ Lorentz frame by introducing the momentum vector $p_{2}^m$ (which appears in the superalgebra (\ref{special})) taken
along {\it only the compactified dimensions} $(c+1,1)$ \cite{sentropy}. The entropy is then an invariant of SO$(c+1,1)$, as it should be, since counting states in representations of the superalgebra could not depend on the basis used to classify them.  However, in the general Lorentz basis, the entropy is expressed in terms of the momentum $p_{2}^m$ and central extensions that are classified by  SO$(c+1,1)$. Similarly, the mass of the black hole is expressed through the same quantities. Thus, such p-brane charged black holes ``know" about the existence of the $(c+1,1)$ dimensions as exhibited in the expressions for the entropy and mass. This is in
agreement with the cosmological view expressed in this paper.

Encouraged by these observations, we propose the following plausible scenario: Before
the Big Bang phase transition the fundamental theory could be a theory of
interacting open and closed super p-branes propagating in a spacetime of
signature $\left( 11,2\right) $ or $\left( 11,3\right).$ The form of the
theory is unknown but it is assumed to obey the general algebra of S-theory. The
interaction of the p-branes is taken such that there are constraints on their
relative motion that effectively forces them to propagate with a single
timelike coordinate (as in the two particle example discussed in section 2,
and generalized in another paper to strings \cite{ibkounnas2}). The constrained system still allows the p-branes to propagate in various linear combinations of the two timelike dimensions as illustrated by the solution $\Phi(x_1,x_2)$ in the two particle example. However, when the Big
Bang phase transition takes place (for reasons that remain obscure at the present) some of the p-branes are associated with
one timelike coordinate which parametrizes the expanding portion of the
universe while others remain in the compactified space that includes the
other timelike coordinate. For low energy physics it is that sector of solution space which becomes relevant. The information about the extra timelike coordinate
would not be lost to observers or matter in the expanding universe because low energy
physics would need to take into account the geometry of the internal space
which would have Minkowski signature.

If one would take this scenario seriously then the superalgebra that would apply in
low energy physics would naturally be along the lines of the ``new
supersymmetry'' discussed in our previous paper \cite{superpv}. For example
in four dimensions 
\begin{equation}
\left\{ Q_{\alpha a},Q_{\dot{\beta}\dot{b}}\right\} =2\sigma _{\alpha \dot{
\beta}}^\mu \gamma _{a\dot{b}}^m\,p_{1\mu }\,p_{2m} + {\rm extensions}, \label{compactified}
\end{equation}
where $p_{1\mu }$ is a vector in SO$\left( 3,1\right)$ corresponding to
ordinary momentum in 4D and $p_{2m}$ is a vector in SO$(c+1,1)$
corresponding to the momentum of the p-branes in the internal dimensions.
The fermionic and bosonic fields that form representations of this
superalgebra depend both on the 4D and internal coordinates $\Phi (x^\mu
,y^m)$ as discussed in our previous work. A Kaluza-Klein type of expansion
in a complete set of states of the internal geometry identifies the low
energy massless families. It was shown in \cite{superpv} that this formalism
can provide a new source of family generation (and unification) that is different than the
standard Kaluza-Klein mechanism. A simple example of how massless families could emerge while being unified in a new supermultiplet was given in \cite{superpv}. The essential difference is that $\gamma _{a\dot{b}}^mp_{2m}$ appears in the form of a product with $\sigma _{\alpha\dot{\beta}}^\mu p_{1\mu }$ in the case of the new supersymmetry, whereas it appears in the form of a sum (and only spacelike components in $p_2$) in the case of standard
supersymmetry. The nature of the internal geometry would need to follow from
more detailed theories, but as is already clear from the simple example, one would not be limited to only Calabi-Yau spaces in spacelike dimensions since the internal space would have Minkowski signature in the new scenario. 

We do not claim to have resolved all issues related to two timelike dimensions in this paper, but we think we have illustrated how to surmount some of the major and obvious obstacles. Besides low energy physics one expects that the extra timelike dimension(s) would have some other consequences, for example in the early universe. Our approach needs to be applied and tested in cosmology. Since the mechanism we have suggested is consistent with and has clear connections to several lines of thought to the unknown theory behind string theory, it would be interesting to explore its consequences and its consistency in more detail by reconsidering those theories.

\end{document}